\documentstyle[a4,umlaute,epsf,12pt]{article}

\newcounter{saveeqn}

\newcommand{\be}{\begin{equation}}
\newcommand{\ee}{\end{equation}}

\newcommand{\bea}{\begin{eqnarray}}
\newcommand{\eea}{\end{eqnarray}}

\begin{document}

\begin{center}

{Submitted to Journal of Rheology}

{\bf \large Characterization of long-chain branching effects in linear rheology}

\vskip 2.0 true mm
\noindent
\sc {Wolfgang Thimm$^{1}$\\
Christian Friedrich$^{1}$\footnote{Corresponding author: 
e-mail: chf@fmf.uni-freiburg.de} \\
Tobias Roths $^{1}$\\
Stefan Trinkle $^{1}$\\
Josef Honerkamp $^{1,2}$} 
\vskip 2.0 true mm
{\it $^{1}$ Freiburger Materialforschungszentrum, Albert-Ludwigs-Universit\"at
Freiburg, Stefan-Meier-Stra{\ss}e 21,
 D-79104 Freiburg im Breisgau, Germany}\\
\vskip 3.0 true mm
{\it $^{2}$ Albert-Ludwigs-Universit\"at Freiburg, Fakult\"at f\"ur Physik, 
Hermann-Herder-Stra{\ss}e~3, D-79104 Freiburg im Breisgau, Germany}

\end{center}

\section*{Synopsis}

This is the last part of a series of five articles published in Journal of Rheology
(Maier et al. (1998), Thimm et al. (1999a), Thimm et al. (2000a), Thimm et al. (2000c)) in which
progress on the determination of binary molecular weight
distributions from rheological data has been reported. In this article is
discussed in how far the developed methods can also be used to characterize
long-chain branching effects.

Monomodal samples which contain long-chain branches
will show
two relaxation processes in the rheological examination, which are converted to two
peaks in a corresponding molecular weight distribution.
But these samples will show only one peak in a molecular weight distribution determined
by a
size-exclusion chromatography examination. This difference can be 
used to characterize long-chain branched materials as will be explained in 
this article.

The usefulness of this method is demonstrated by examining
polymers, which contain definite long-chain branches specified from
the way, they were made.

\section*{I. Introduction}

Thimm et al. (1999a, 1999b, 2000a, 2000b, 2000c) have shown recently,
how the molecular weight distribution (MWD) 
of binary polymer blends made of monodisperse 
components of linear polymers
can be reconstructed from rheological data. The quality of the reconstruction
gave results comparable to a size-exclusion chromatography (SEC) determination of 
the molecular weight distribution, which is a well-established method.

A central key of their method is the determination of the relaxation time
spectrum. 
While the relationship between the molecular weight distribution of linear
polymers and the linear rheological properties is henceforth well understood,
there is another problem in polymer characterization,
where rheology might be useful. This is the examination of
the relationship between long-chain branching effects and rheological
properties. So far there is no method available to characterize
long-chain branching in a robust and reliable manner.
In this article we discuss how the relaxation time spectrum and
the method developed by Thimm et al. (1999a, 2000a)
can be used examining this problem.

For polymers with long-chain branches (armes), which are longer than the entanglement
molecular weight ($M_e$), it is found, that the terminal relaxation
time shifts with the arm length towards longer times.
This shift corresponds to violation of
the well-known 3.4 scaling law between molecular weight ($M_w$) 
and the zero shear rate viscosity
$\eta_{0}$.

Furthermore, the relaxation time spectrum of long-chain
branched materials shows two peaks corresponding to two relaxation
processes. 
These can be interpreted as the processes corresponding to
the relaxation of the arms and the relaxation of the whole molecule.

To clarify our notation, we name the molecular weight distribution, which was determined
with size-exclusion chromatography 'size-exclusion chromatography
determined molecular weight distribution' (sMWD),
while the molecular weight distribution determined from rheological data is named 
'rheological molecular weight distribution' (rMWD). The difference
between rMWD and sMWD will be called
'difference molecular weight distribution' (dMWD). In this article will be discussed
that this dMWD can be reasonably interpreted as molecular weight distribution of the
branches (arms).

While the rheological relaxation processes are well reflected in the
relaxation time spectrum, the size-exclusion chromatography can not
differentiate these processes. Therefore, there is just one peak in
a size-exclusion chromatography determination of the sMWD from a
monomodale
long-chain branched polymer sample. On the other hand
one finds a binary distribution, in the rMWD,
since there are two relaxation processes. These differences therefore report, whether a
sample contains long-chain branches or not.

The two peaks in the rheological rMWD are explained on a phenomenological level in
this article.
We take the picture as guideline that the shorter relaxation processes
in the relaxtion time spectra correspond to the relaxation of the
arms, while the longer relaxation processes correspond to the relaxation
of the whole polymers.
Then it seems plausible that the
shorter relaxation processes, when converted to a molecular weight
distribution, tell the polydispersity and length of the branches,
while the peak at higher relaxation times corresponds to the relaxation
of the polymer. It is plausible that the peak might be found at higher
molecular weight than would be determined with the size-exclusion
chromatography, since the branches prevent reptation and this can
lead to slower relaxation.

The method itself is described in section II.
In section III. results found for specially designed H-shaped molecules
are discussed. In section IV. we discuss, in how far these results can be transferred
to polyolefines. In section V. the conclusions are given. Finally, in the
appendix the relation between the
results obtained using the novel method to established theoretical work on long-chain
branching in polymers is discussed.

In the remaining subsection of this introduction we briefly reflect on the
method introduced by Thimm et al. (1999a, 2000a) to convert the relaxation time
spectrum into a rheological molecular weight distribution.

\subsection*{Relaxation time spectrum and molecular weight distribution}

Thimm et al. (1999a) derived some analytical relations, which relate the
relaxation time spectrum $h(\tau)=\tilde h(m)$ (with $\tau=\tau(m)$)
to the molecular weight distribution $w(m)$ for linear polymers.
The rheological molecular weight distribution is given by:
\be \label{rel} w(m) =  \frac{1}{\beta}\frac{{\alpha}^{(1/\beta)}}
{(G_N^{0})^{1/\beta}
} \tilde h(m)
 (\int_{ m}^{\infty} \frac{\tilde h(m')}{m'} 
{\rm d} m')^{(1/\beta-1)},\ee
and the inverse relation is given by:
\be \label{inverse}
\frac{\tilde h(m)}
{G_N^0}=\frac{\beta}{\alpha} w(m) [\int_m^\infty {\rm d} m' 
\frac{ w(m')}{m'}]^{\beta-1}
.\ee
In these equations
the generalized mixing parameter is about two ($\beta=2$), when the Rouse
spectrum is treated 
separately in the data evaluation (Thimm et al. (2000a)). The plateau modulus is denoted $G_N^0$.
On the one hand these relationships are especially useful to understand rheological
data from binary molecular weight distributions. On the other hand,
this knowledge can be used to compare the results of rheological measurements
to these obtained with size-exclusion chromatography.
However in addition to the conventional determined rheological data (dynamic shear
moduli, creep function) the constants $k, \alpha \approx 3.4$ in the scaling relation 
\be \label{tauref}
\tau=km^{\alpha},
\ee
where $\tau$ is a relaxation time and $m$ a normalized molecular weight,
have to be determined experimentally.

\section*{II. The method}

In this section the novel method is introduced.
To use this method, the following quantities have to be determined:

\begin{itemize}
\item[(i)] The relaxation time spectrum $h(\tau)$ has to be estimated
from measured rheological data
(e.g. shifted dynamic moduli $G'(\omega)$, $G''(\omega)$).
(How this estimate can be obtained is
described e.g. in Roths et al. (2000).)

\item[(ii)]
The 'size-exclusion chromatography determined molecular weight distribution' should be
measured
experimentally.
\end{itemize}

We make the following assumptions (if the parameters are not
known from different experimental determinations) concerning
the parameters, which are needed to calculate the molecular weight
distribution.

\begin{itemize}
\item[(iii)]
For the value of the generalized mixing parameter $\beta$ we set the
theoretical value $\beta=2$.

\item[(iv)]
For the scaling parameter $\alpha$ we set the universal value
$\alpha=3.4$.

\end{itemize}

With these parameters fixed there are two additional 'free' parameters left,
which have still to be determined. This determination is 
an essential non-trivial step in this method.

\begin{itemize}

\item[(v)]
The time where the Rouse spectrum vanishes (named $\tau_R$ in Thimm et al. (2000a))
can be accurately determined using the following idea:

When $\tau_R$ was determined wrongly, there would be a discontinuity in the
rheologically determined molecular weight distribution at the
position corresponding to $\tau_R$. Therefore, we take the value for $\tau_R$,
which guarantees a smooth rheological molecular weight distribution in the region
between the determined $\tau(m_e)$ and the terminal relaxation time.

\item[(vi)]
A well-established observation is that
the value $k$ in the given scaling relation Eq. (\ref{tauref}) depends strongly on the
temperature and the molecular weight of the branches (arms)
(see appendix).

Therefore, it is not possible to use constant literature values
or values determined from linear polymers to determine $k$. The result of such a use
would be that the peak corresponding to the sMWD would be found at molecular weights, which are
too high. Also the use of a different theoretical scaling relation would be of little help, since
for a branched polymer sample of unknown topology the details in these relations would differ
strongly.

To be able to reasonably combine sMWD and rMWD,
we suggest to determine the terminal relaxation
time $\tau_{term}$, which shows as terminal relaxation peak in the
relaxation time spectrum. The position of this
peak is for linear polymers related to the maximum in the SEC determined molecular
weight distribution $m_{term}$. When both parameters are well determined,
it is possible (inserting both values in Eq. (\ref{tauref})) to calculate
$k$ accurately.

The $k$, which is determined this way is not the microscopic $k$, describing the properties of
the tube in the reptation picture (Doi and Edwards (1986)), discussed above.
However, as will be discussed below, the typical error, which will be made using this procedure, is
acceptable and the procedure will give reasonable values for the arm molecular weight $M_{arm}$.

The practical realization of this thought could be to calculate in a first step the
rMWD with an essentially arbitrary $k$ (for example the $k$ of linear polystyrene). In a second
step the rMWD could be scaled such that the peak with the higher molecular weight matches
the peak in the sMWD.

\end{itemize}

With the steps i) to vi) all is known needed to determine the
rheological molecular weight
distribution from the estimated relaxation time spectrum -
using the procedure given in Thimm et al. (1999a, 2000a).

As outlined in the introduction, for a monomodal sMWD
the rMWD is also monomodal
for a sample without long-chain branches, but is binary
for a sample containing long-chain branches.

The peak corresponding to the higher molecular weight is related to the relaxation
of the whole molecule. 

\begin{itemize}

\item[(vii)]
Therefore, we
drop the peak corresponding to the higher molecular weight.
This drop corresponds in general to build the difference between the
rheologically determined molecular weight distribution with the
molecular weight distribution determined with SEC.

\item[(viii)]
The peak corresponding to the lower molecular weight reflects the molecular
weight distribution of the long-chain branches (arms) and therefore
characterizes long-chain branching effects in polymers.

\end{itemize}

A parameter, which could be used for quantitative comparison, could be the
average of the dMWD. We name this average $M_{arm}$ and discuss in the
following subsection, that this average will be of the same order, as would be expected for
the microscopic $M_{arm}$.

\subsection*{Typical values}

To give a feeling for typical values for $M_{arm}$, which can be expected
using this method, we discuss some simple considerations
in this subsection.
We are interested in typical values, which are expected for
$M_{arm}$, when $M_{term}$ is given by determination of sMWD.

The most simple argument is that obviously the following
inequality must hold:
\be
M_e < M_{arm} < M_{term}.
\ee
From topological considerations (the 3 arm star is the most
simple branched molecule) it is clear that 
$M_{arm}<1/3M_{term}$. Typically the reptation starts at
the critical molecular weight $M_c$, which is about
a factor 2 greater than the entanglement molecular weight.
To show effects the arm molecular weight should therefore
be at least of this order $2M_e<M_{arm}$.
When we consider the value for $M_{term}$ of $M_{term}
\approx {\rm 1. Mio. {}} g/mol$,
and the typical value for the entanglement molecular weight
$M_e \approx 10000 g/mol$, we find approximately that:
\be
1/50 M_{term}< M_{arm} < 1/3 M_{term}.
\ee
So a determination of the arm molecular weight using 
rheological means by given $M_{term}$ is obviously 
restricted to arms length within one
decade.

Another consideration is that one may insert the scaling
relations Eq. (\ref{tauref})  for $M_{term}$ and
$M_{arm}$ into each other. One finds easily:
\be
M_{arm}=(\frac{\tau_{arm}}{\tau_{term}})^{1/\alpha}M_{term}.
\ee
We take for the ratio of $\tau_{arm}$ and
$\tau_{term}$ the typical values $100$ to $10^6$.
The value $100$ should be assumed to distinguish binary
behaviour and the value $10^6$ represents the typical
limits of the frequency window, which is accessible using
rheology. 
When we insert these values together with the
assumed scaling parameter $\alpha \approx 3.4$ we find
the approximate values:
\be
1/60 M_{term}< M_{arm} < 1/4 M_{term}.
\ee
Both appraisements give for $M_{arm}$ a typical
value of about $1/10 M_{term}$.

\section*{III. Data evaluation}

In this section data from well-defined H-shaped polymers are examined.
The data from the shifted dynamic moduli $G'(\omega)$,
$G''(\omega)$ used, are given and discussed in (Roovers (1984), McLeish et al. (1999)).
Moreover the original SEC-curves, provided generously by J. Roovers, were checked to
confirm the result that the SEC-curves showed indeed just one peak.

\subsection*{A. Polystyrene}

In this section we discuss results found for specially designed H-shaped polystyrene
(Roovers (1984)).  

Roovers (1984) has constructed three samples of perfectly designed H-shaped 
molecules where the molecular weight of the arms 
and the bar are the same. The ones we discuss in this article are named
H2A1 ($M_w=2.37 \cdot 10^{5} g/mol$), H1A1 ($M_w=4.83 \cdot
10^{5} g/mol$) and H5A1 ($M_w=10.4 \cdot
10^{5} g/mol$). The weights for the arms can be obtained by division $M_w/5$.
These were examined rheologically and later two 
groups have found that these rheological data
contained indeed two relaxation processes (Friedrich et al. (1995),
Hatzikiriakos et al. (2000)).
The first group used a very stable analytical ansatz designed to find two relaxation 
processes in rheological data, while the second group found the analogous result using a
discrete relaxation time spectrum.

The scaling parameters of linear polystyrene are determined experimentally
(Maier et al. (1998)) to
$k=6.919 \cdot 10^{-20}$ sec ($m$ given in $g/mol$) and $\alpha=3.67$. We use these
parameters, having in mind that the peaks of the averaged molecular weights and the
relaxation time spectrum need not match in our evaluation.
We calculate the relaxation time spectrum and determine the rheological
molecular weight distributions.
The results are shown in figures 1a)-c)
together with the averaged molecular weight of the nearly monodisperse polystyrene given
by Roovers (1984) for the whole molecule and the arms.

The results show that the idea that the molecular weights of the
molecules and the arms are reflected well by the rheological determined molecular
weight distribution (rMWD) seems plausible. The shift seems to become large just when
the mass of the arms exceeds the entanglement molecular weight, which is
known to be for polystyrene about $M_e=18000$ g/mol.
The agreement between rheologically determined weights
and the weights given from SEC by Roovers is remarkable.
It seems plausible that the polydispersity given in the figures reflects 
the polydispersity of the linear polystyrene, which was used to construct the samples.

\subsection*{B. Polyisoprene}

A theory, which describes the dynamics of H-polymers in greater detail
was worked out and discussed by McLeish et al. (1999). However
this group did not solve the ill-posed inverse problem to estimate the
spectrum from the measured dynamic moduli.

McLeish et al. (1999) have examined and published data (in Fig. 6 of
the cited article) of four H-polymers
from polyisoprene.
The samples are named:\\
H110B20A ($M_w(SEC)=198000 g/mol, M_{arm}(SEC)=20000g/mol$), \\
H160B40A ($M_w(SEC)=324000 g/mol, M_{arm}(SEC)=40000g/mol$), \\
H110B52A ($M_w(SEC)=310000 g/mol, M_{arm}(SEC)=52500g/mol$), \\
H200B65A ($M_w(SEC)=460000 g/mol, M_{arm}(SEC)=63000g/mol$).\\
 The results for the rheological molecular weight distributions
found with the 'step in' parameters for polystyrene are given in Fig.'s 2a)-d).
The binary behaviour can be clearly identified.
When the constant $k$ is determined as described above,
the values are given in McLeish et al. (1999),
the molecular weight distribution of the arms can be calculated with
reasonable values comparing to the 
molecular weight averages for the arms given by McLeish et al. (1999).
Results from the sample with the highest arm molecular weight are shown
in Fig. 3.

We see in Fig. 2d), that it is important to check the validity of the
procedure, that the frequency window, which is used to determine
the rMWD, should be wide enough to contain two clearly separable peaks.

The agreement between the $M_{arm}$ given by construction and the $M_{arm}$ determined with
the procedure proposed in this article seems to become better for the higher arm molecular weight.
The same is observed for the polystyrene data.

A possible explanation for this observation could be that for the lower arm molecular weight samples
$M_{arm}$ is too close to the entanglement molecular weight $M_e$, to be well resolved.
Below $M_e$, we assume the relaxation time spectrum to be dominated by the Rouse modes, while the
influence of $M_{arm}$ should be vanishing in this region. For this reason it is also
expected that the
relaxation time spectrum should not reflect the microscopic molecular weight distribution of
the arm molecular weight below $M_e$ correctly.

\section*{IV. Possible limitations concerning polyolefines}

In this section results for an evaluation for various polyolefine samples
(linear and branched) are summarized.
(See also Friedrich et al. (1999), 
a more detailed evaluation will be published elsewhere.)
Examining rheological data from polyolefines,
we found binary behaviour of the rMWD in one sample (See Fig. 4).
Taken the value of $k$ for polystyrene realistic values for $M_{term}$
and $M_{arm}$ are obtained.
The rMWD shows indeed a broad peak in the rMWD, which can be interpreted as 
molecular weight distribution of the arms.
The position of the peak is in good agreement with the theoretical considerations discussed in 
section II.

In general it is found that the binary behaviour is less clearly detectable than with 
the monodisperse H-shaped molecules of Roovers (1984). 
This observation can be motivated as follows:
\begin{itemize}
\item[(i)] While the polystyrene samples contain nearly monodisperse ($M_{w}/M_{n} \approx
1.02$)  parts, metallocene catalysts produce broader distributed polymers
($M_{w}/M_{n} \approx 2$). The broadening of the samples can be a reason that the
binary behaviour is smeared out.
\item[(ii)] Since it is unclear how the long-chain branches in the samples are really
distributed, it might be that the peak corresponding to the arm molecular weights
is smeared out considerably, since the long-chain branches have very different
lengths and positions.
\item[(iii)] The frequency windows of the (from dynamic moduli and creep curve constructed)
mastercurves is limited. This limitation limits the mass range
in which the rMWD can be accurately determined considerably
as has been reported by Thimm et al. (2000b) for linear high density polyethylene.
Therefore, the peaks of the corresponding binary molecular weight
distributions may be outside the rheologically acessible frequency window.
\end{itemize}

Although the parameters for $k$ and $\alpha$ are essentially unknown for the
new materials under consideration, the
method described in this article can still give interesting information without essential
limitations. The points discussed above are of technical manner, which
are in principle not unsolvable. Therefore, these points
do not question the idea of our novel approach.

\section*{V. Conclusion}

For linear polymers the relationships between the molecular weight
distribution and rheological data
have been well examined (Thimm et al. (1999b, 2000a)).
When the developed procedure is applied to branched polymers,
we find that the rheologically determined molecular weight distribution
can be reasonably interpreted as a reflection of the
molecular weight distributions of the whole molecules and of the arms of
the branched molecules.

Starting from this idea we have developed a novel method, which allows
robust characterization of long-chain branching in polymers.
This method is motivated on a phenomenological level.

Using this method, it is possible to explicitly determine
the molecular weight distribution of the branches (arms), which is
not accessible so far by any other method.
We therefore think
this method to be a valueable tool to characterize long-chain
branching effects in polymers.

\section*{Acknowledgements}

S.Trinkle was supported by the Deutsche Forschungsgemeinschaft:
Graduiertenkolleg f\"ur Strukturbildung in makromolekularen Systemen.

\section*{Appendix: Connection to established theories}

The influence of the quantity 'arm molecular weight' $M_{arm}$
on various rheological quantities (like the zero-shear rate
viscosity $\eta_0$, or the terminal relaxation time $\tau_{term}$) has been
studied intensively in the literature (see e.g. Doi and Edwards (1986)).
A well-known relationship
(see e.g. Doi and Kuzuu (1980))
is the relation between the terminal
relaxation time $\tau_{term}$ and the arm molecular weight $M_{arm}$ relative
to the entanglement molecular weight $M_e$ and the corresponding relaxation time
$\tau(m_e)=\tau_e$:
\be
\tau_{term}=\tau_e(\frac{M_{arm}}{M_e})^b \exp(\nu\frac{M_{arm}}{M_e} ),
\ee
where in good approximation one can take $\tau_{term}=\eta_0/G_N^0$.

Once, $M_{arm}$ is determined with the new method ($M_{arm}$ is the average molecular
weight of the arm molecular weight distribution),
it is straightforward to insert $M_{arm}$ and these
rheological quantities in the
relations from literature. Solving such equations it is possible
to estimate additional parameters (e.g. $b$ for given $\nu$), which are relevant in
the theory of branched polymers.

\section*{References}

\noindent
Doi, M. and S. F. Edwards, {\it The Theory of Polymer Dynamics},
(Clarendon, Oxford 1986). \\

\noindent
Doi, M. and N.Y. Kuzuu,
``Rheology of star polymers in concentrated solutions and melts'',
J. Polym. Sci. Polym. Letters. Edn.
18, 775-780,
 (1980). \\

\noindent
Friedrich, C., H. Braun, and J. Weese ``
Determination of Relaxation Time Spectra by Analytical Inversion 
Using a Linear Viscoelastic Model with 
Fractional Derivatives'', 
Polym. Eng. and Sci. 35, 1661-1669, (1995). \\

\noindent
Friedrich, C.,  A. Eckstein, F. Stricker, R. M\"ulhaupt, 
``Rheology and Processing of metallocene-based polyolefins'',
in J. Scheirs (Ed.) 
{\it Preparation, Properties and Technology of Metallocene-Based 
Polyolefins}, John Wiley and Sons, (1999) \\

\noindent
Hatzikiriakos, S. G., M. Kapnistos, D. Vlassopoulos, 
C. Chevillard, H. H. Winter, J. Roovers:
      `` Relaxation time spectra of star polymers '',
        Rheol. Acta 39 (1) , 38-43, (2000).\\

\noindent
McLeish, T.C.B., J. Allgaier, D.K. Bick, G. Bishko, P. Biswas, R. Blackwell,
B. Blottiere, N. Clarke, B. Gibbs, D.J. Groves, A. Hakiki, R.K. Heenan,
J.M. Johnson, R. Kant, D.J. Read, R.N. Young, ``
Dynamics of Entangled H-Polymers: Theory, Rheology and Neutron-Scattering'',
Macromolecules (1999), {\bf 32}, 6734-6758. \\
       
\noindent
Maier, D., A. Eckstein, C. Friedrich, and J. Honerkamp, `` 
Evaluation of Models combining Rheological Data with Molecular Weight
Distribution'',
J. Rheol. {\bf 42 (5)} 1153-1173 (1998) \\

\noindent
Roovers, J. ``Melt Rheology of H-Shaped Polystyrenes'',
Macromolecules (1984), {\bf  17}, 1196-1200\\

\noindent
Roths, T., D. Maier, Chr. Friedrich, M. Marth, and J. Honerkamp, 
`` Determination of the relaxation time spectrum from dynamic moduli 
using an edge preserving regularization method'', 
      Rheol. Acta 39 (2) 163-173 (2000) \\

\noindent
Thimm, W., C. Friedrich, M. Marth, J. Honerkamp,
``An Analytical Relation between Relaxation Time Spectrum and
Molecular Weight Distribution,''
J. Rheol. {\bf 43 (6)}, 1663-1672 (1999a) \\

\noindent
Thimm, W., C. Friedrich, D. Maier, M. Marth, J. Honerkamp,
``Determination of molecular weight distributions from
rheological data - an application to polystyrene, polymethylmethacrylate
and isotactic polypropylene,''
Applied Rheology {\bf 9(4)}, 150-157 (1999b) \\

      \noindent
Thimm, W., C. Friedrich, M. Marth, J. Honerkamp,
``On the Rouse spectrum and the determination of the molecular
weight distribution,''
J. Rheol. {\bf 44 (2)}, 429-438 (2000a) \\

 \noindent
Thimm, W., C. Friedrich, M. Marth, J. Honerkamp,
``Determination of the molecular weight distribution from the relaxation time spectrum,''
 Proceedings of the XIII. International Congress on Rheology,
Binding, Hudson, Mewis, Piau, Petrie, Townsend, Wagner and Walters (Eds.), (1);
107-108 (2000b)\\

\noindent
Thimm, W., C. Friedrich, T. Roths, J. Honerkamp,
``Molecular weight distribution dependent kernels in generalized mixing rules,''
J. Rheol. {\bf 44 (6)}, (in press)(2000c)\\

\unitlength=1 cm

\newpage
{\large \bf FIG. 1a}\\
\unitlength=1 cm
The rheologically determined molecular weight distribution for 
a H-shaped polystyrene sample (H2A1 in Roovers (1984))
. (Step-in parameters of linear polystyrene used). \\
\begin{picture}(1,10.5)
  \epsfxsize=6.7cm
   \put(0,1.7){\epsffile{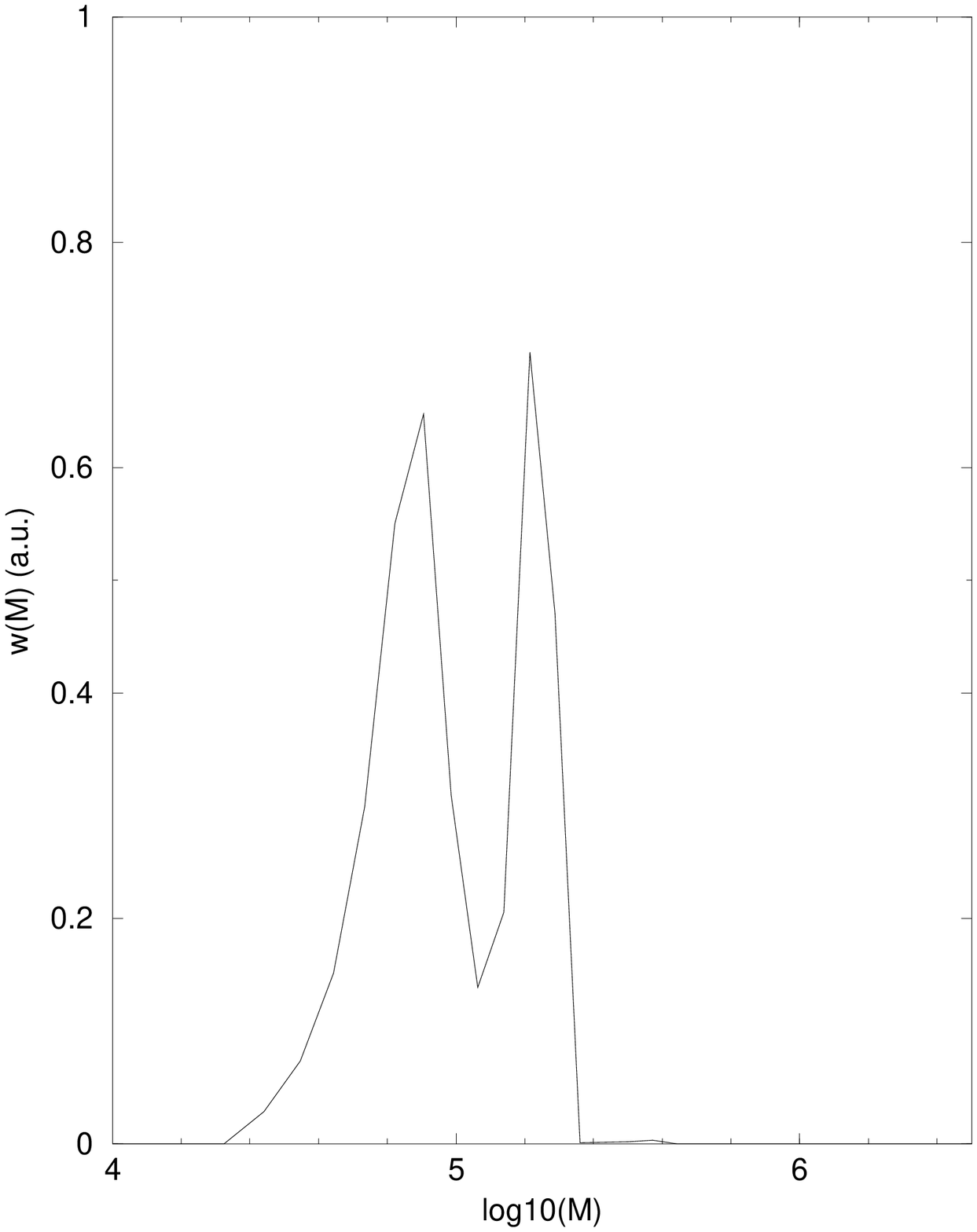}}
 \end{picture}

\newpage
{\large \bf FIG. 1b}\\
The rheologically determined molecular weight distribution for 
a H-shaped polystyrene sample (H1A1 in Roovers (1984))
. \\
\begin{picture}(1,10.5)
  \epsfxsize=6.7cm
   \put(0,1.7){\epsffile{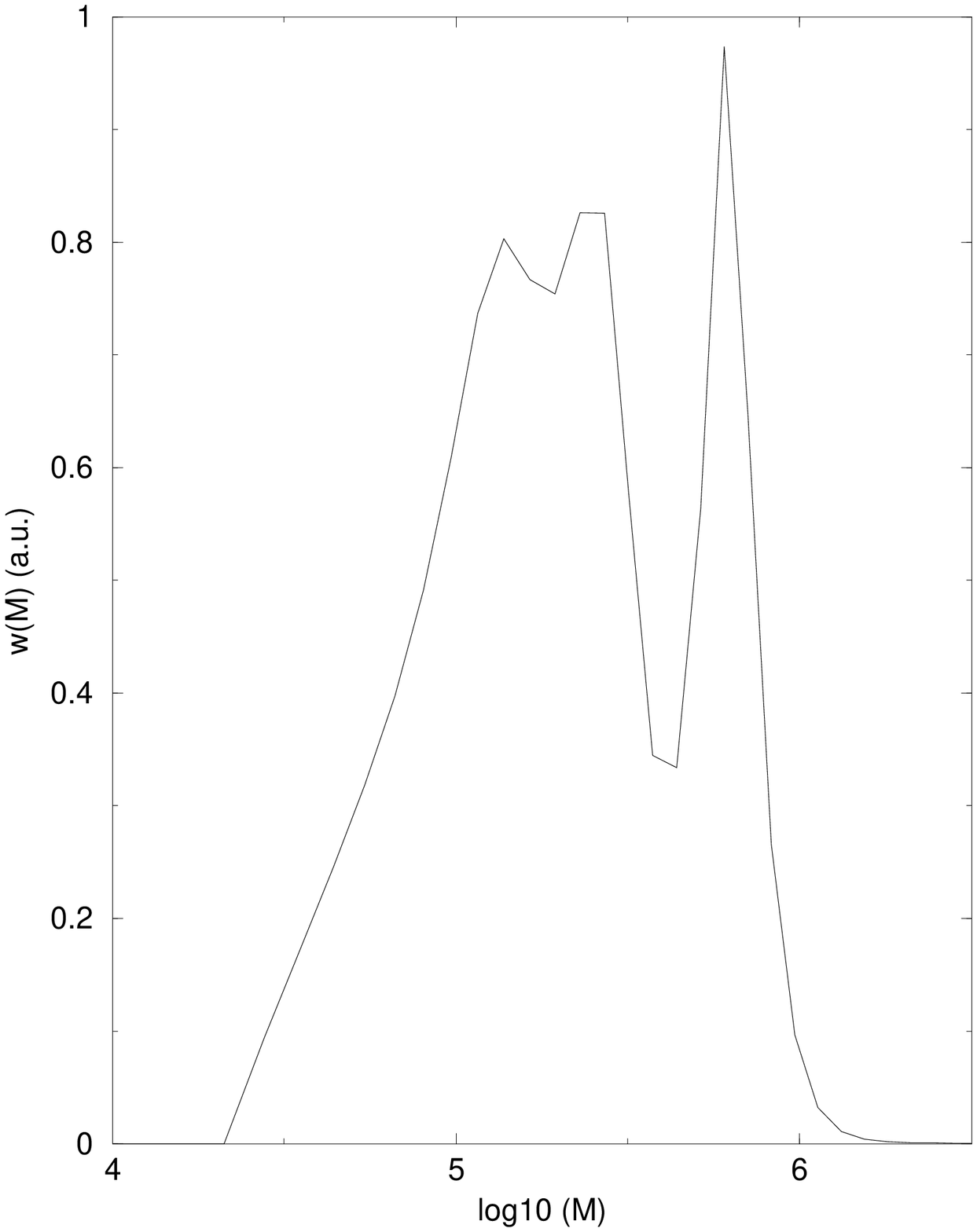}}
 \end{picture}

\unitlength=1 cm

\newpage
{\large \bf FIG. 1c}\\
The rheologically determined molecular weight distribution for 
a H-shaped polystyrene sample (H5A1 in Roovers (1984))
.\\
\begin{picture}(1,10.5)
  \epsfxsize=6.7cm
   \put(0,1.7){\epsffile{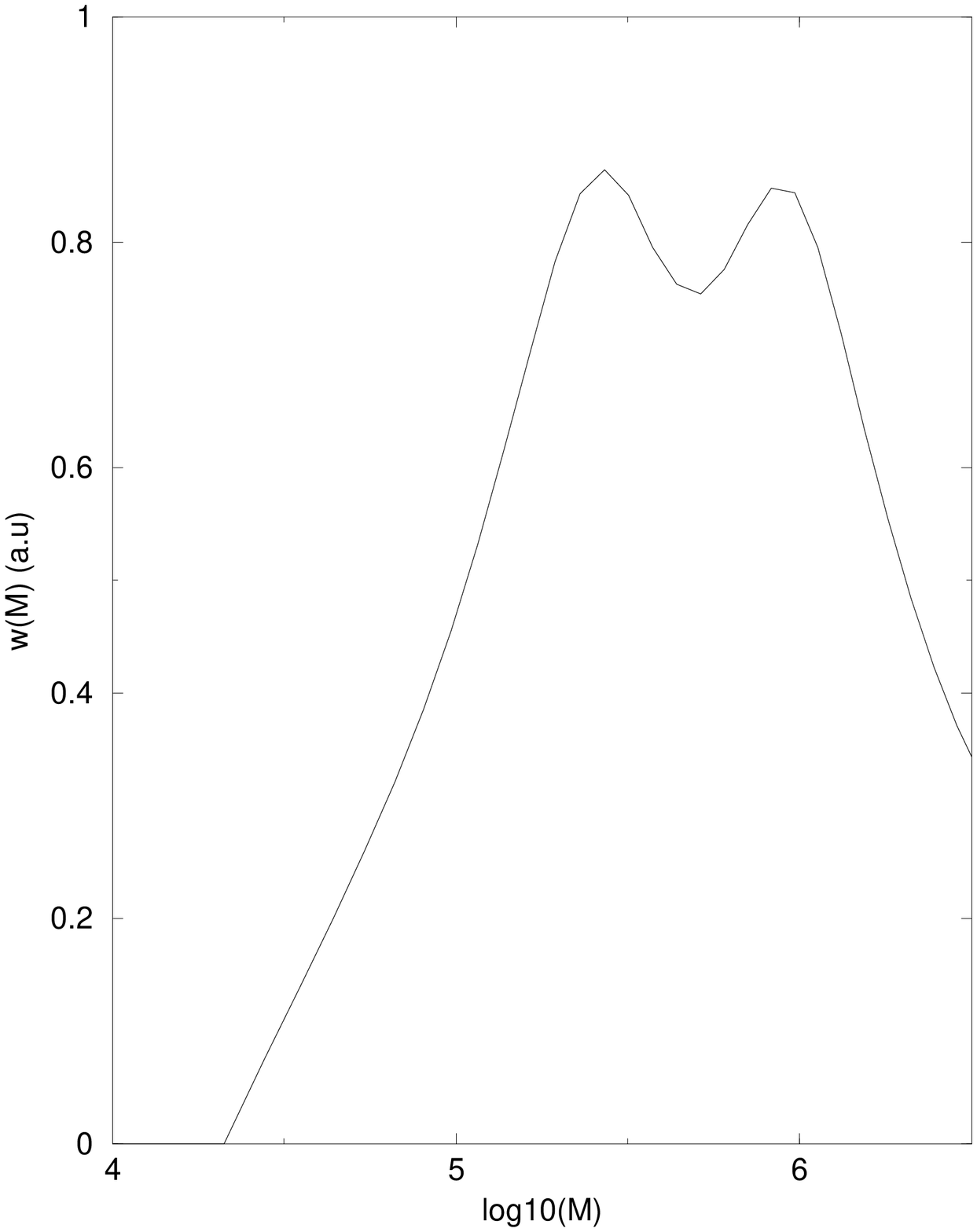}}
 \end{picture}

\newpage
{\large \bf FIG. 2a}\\
The rheologically determined molecular weight distribution for 
a H-shaped polyisoprene sample. $G'(\omega)$, $G''(\omega)$ data taken from
McLeish et al. (1999) (H110B20).
\\
\begin{picture}(1,10.5)
  \epsfxsize=6.7cm
   \put(0,1.7){\epsffile{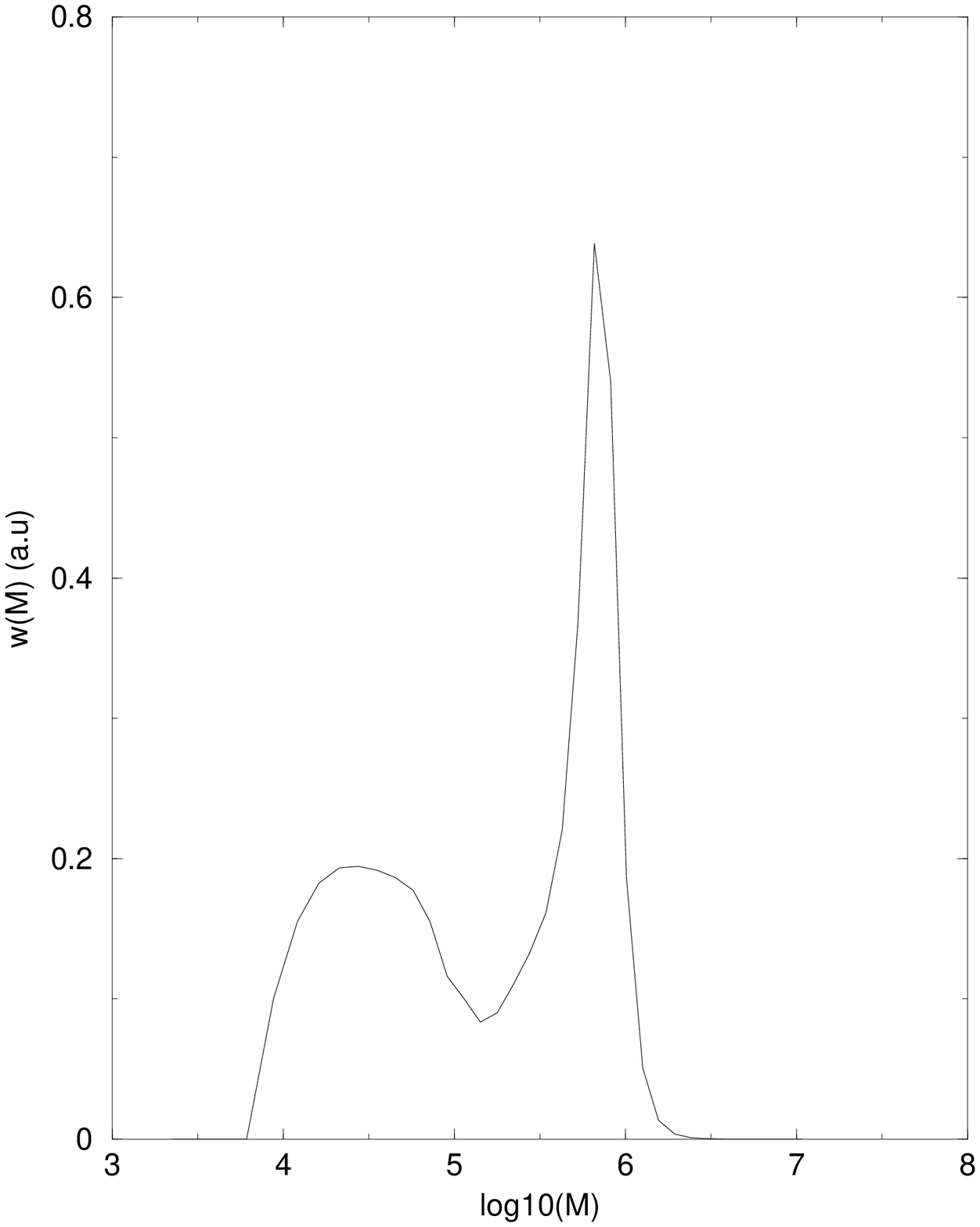}}
 \end{picture}

\newpage
{\large \bf FIG. 2b}\\
The rheologically determined molecular weight distribution for 
a H-shaped polyisoprene sample. $G'(\omega)$, $G''(\omega)$ data taken from McLeish et al. (1999)
(H160B40A).
\\
\begin{picture}(1,10.5)
  \epsfxsize=6.7cm
   \put(0,1.7){\epsffile{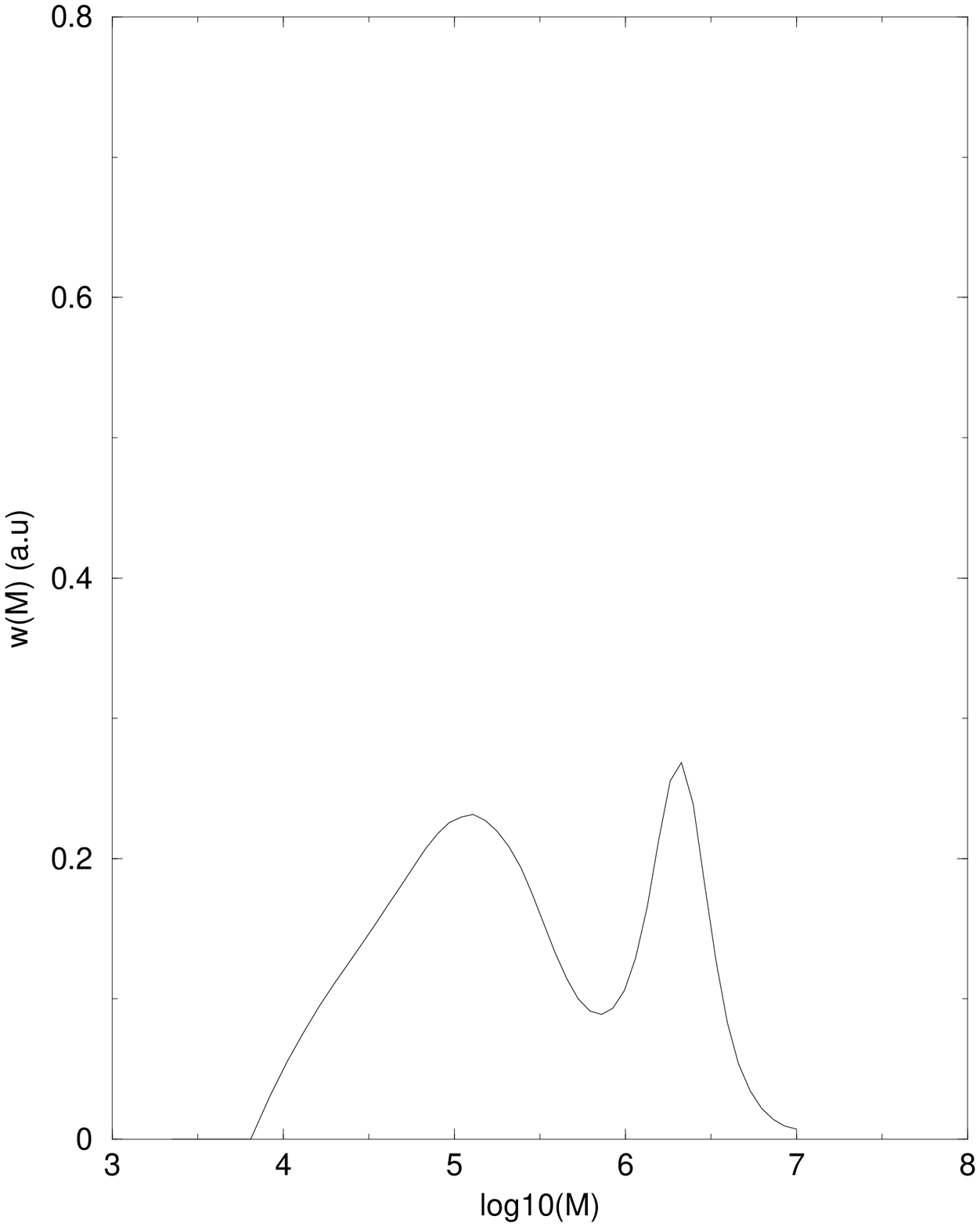}}
 \end{picture}

\newpage
{\large \bf FIG. 2c}\\
The rheologically determined molecular weight distribution for 
a H-shaped polyisoprene sample. $G'(\omega)$, $G''(\omega)$ data taken from McLeish et al. (1999)
(H110B52A).
\\
\begin{picture}(1,10.5)
  \epsfxsize=6.7cm
   \put(0,1.7){\epsffile{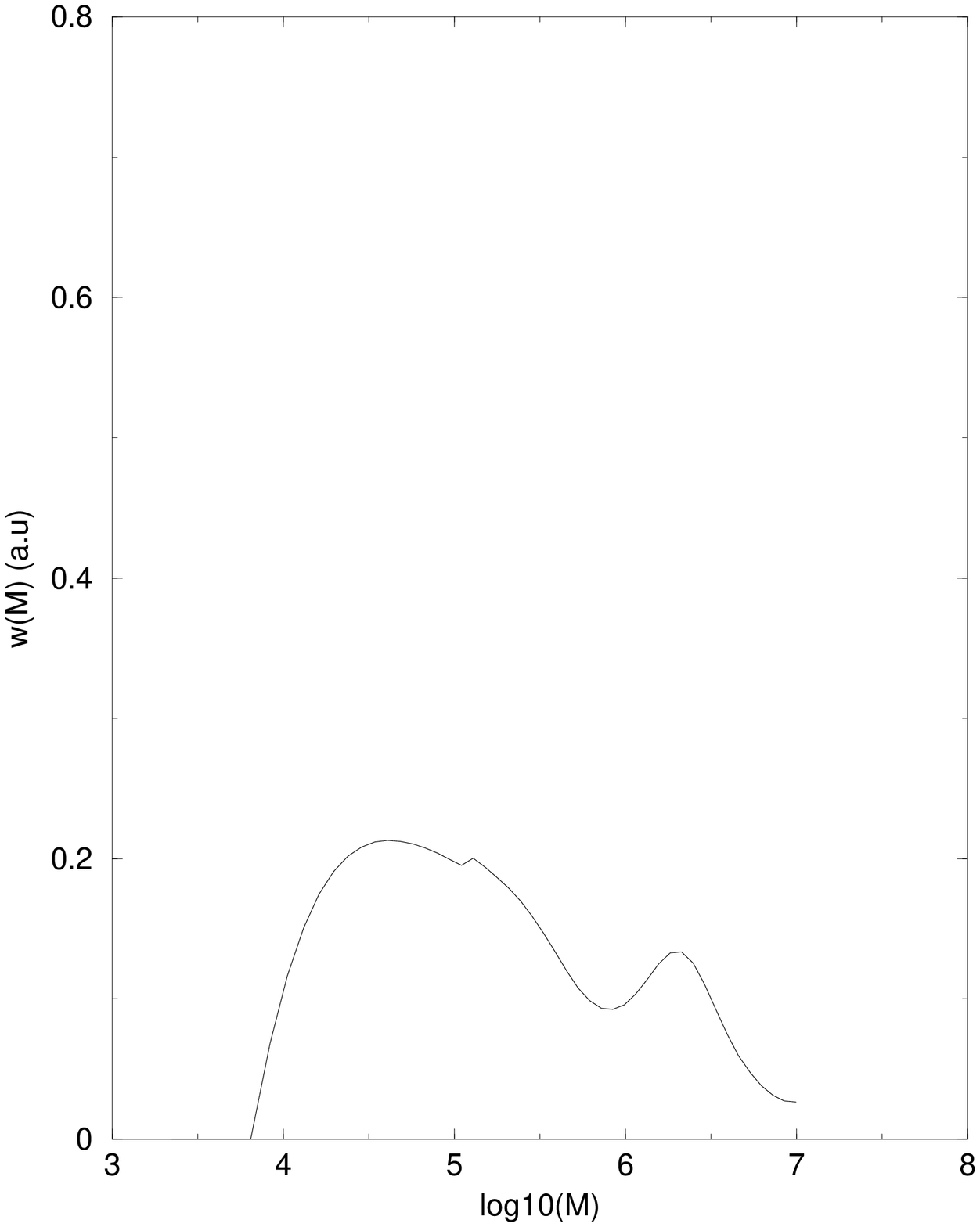}}
 \end{picture}

\newpage
{\large \bf FIG. 2d}\\
The rheologically determined molecular weight distribution for 
a H-shaped polyisoprene sample. $G'(\omega)$, $G''(\omega)$ data taken from McLeish et al. (1999)
(H200H65A).
The peak corresponding to higher molecular weight is incomplete due
to limitations of the frequency range in the measured $G'(\omega)$, $G''(\omega)$
data.\\
\begin{picture}(1,10.5)
  \epsfxsize=6.7cm
   \put(0,1.7){\epsffile{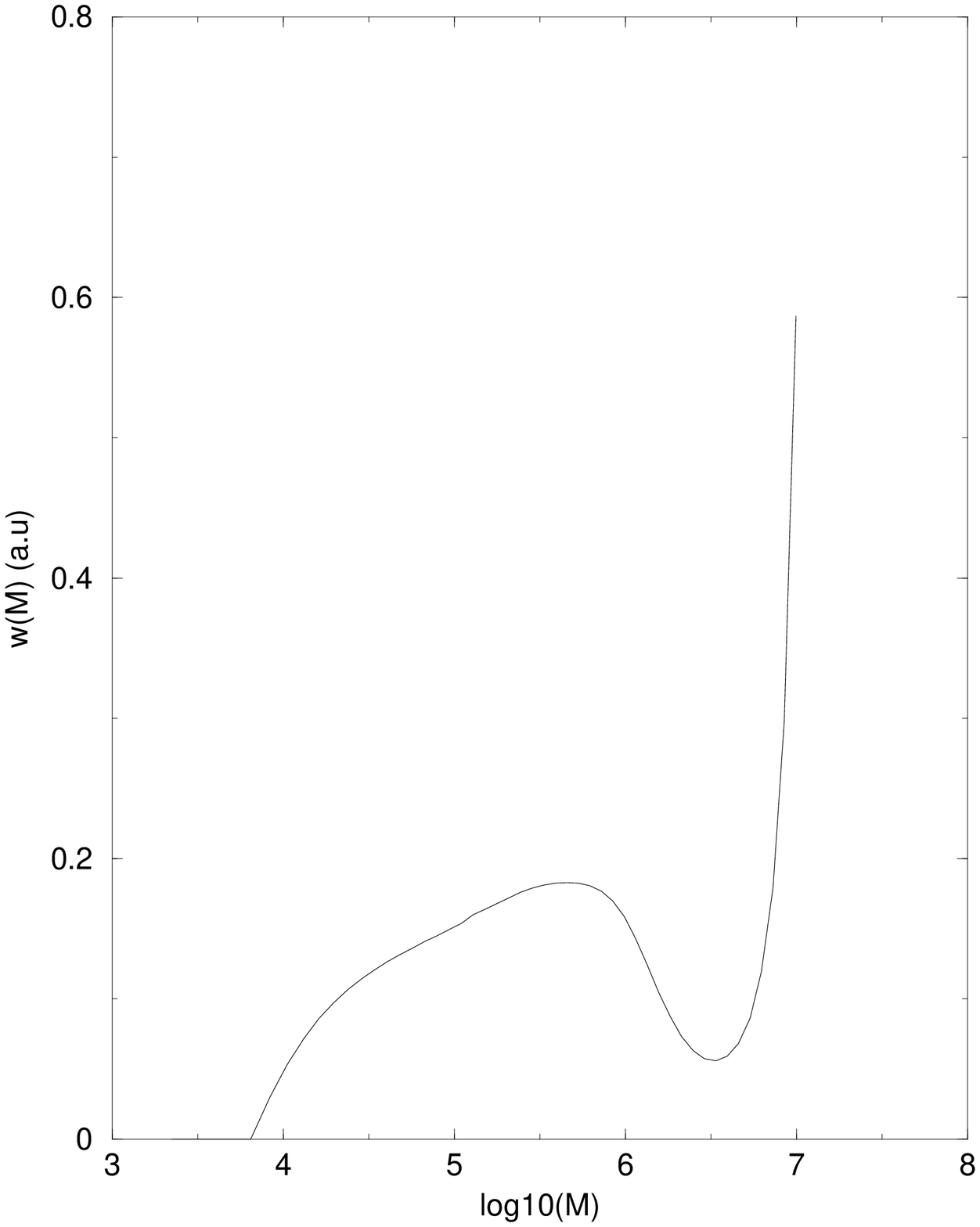}}
 \end{picture}

\newpage
{\large \bf FIG. 3}\\
The distribution of the arm molecular weight's can be clearly
identified using our novel procedure. The value for the average
arm molecular weight is given by McLeish et al. (1999) as about
63000 g/mol.\\
\begin{picture}(1,10.5)
  \epsfxsize=6.7cm
   \put(0,1.7){\epsffile{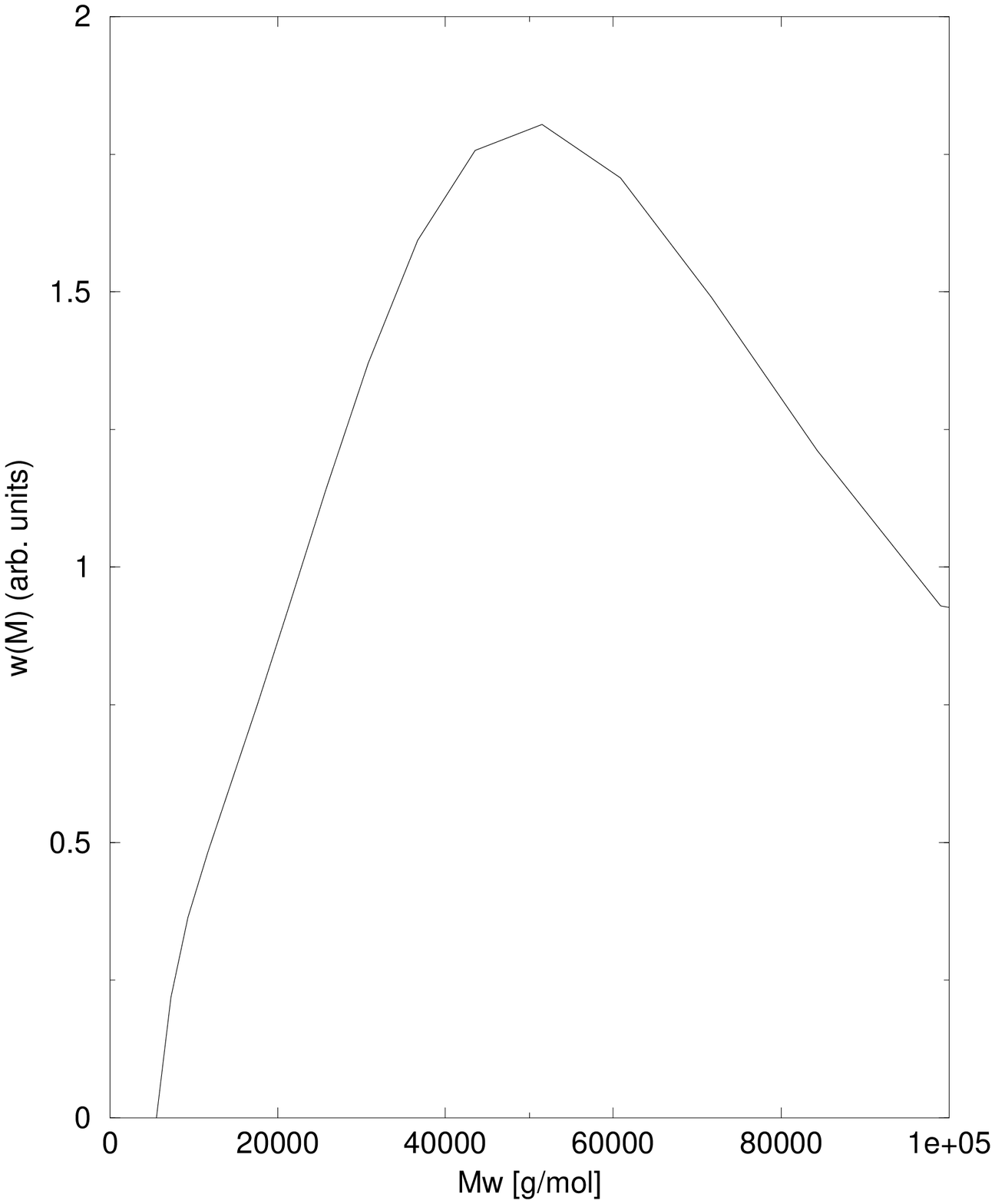}}
 \end{picture}

\newpage
{\large \bf FIG. 4}\\
Examining data from polyolefines, we found a sample,
which showed clearly binary behaviour. \\
\begin{picture}(1,10.5)
  \epsfxsize=6.7cm
   \put(0,1.7){\epsffile{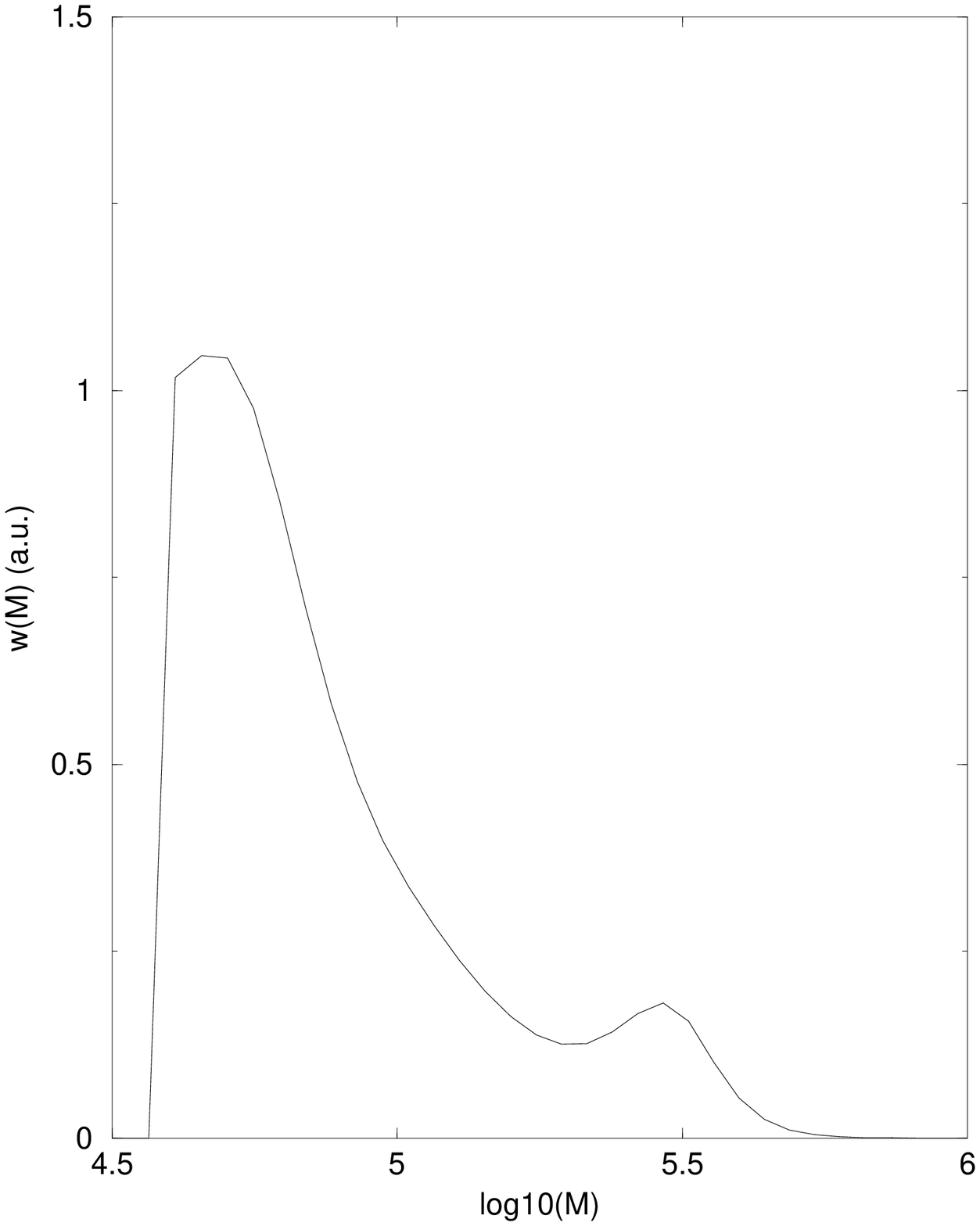}}
 \end{picture}

\end{document}